\documentclass[twocolumn]{aastex61}
\received{... ..., 2017}
\revised{... ..., 2017}
\accepted{... ..., 2017}
\submitjournal{ApJ}

\shorttitle{Formation of cool and hot components in a blowout jet}
\shortauthors{Shen et al.}
\newcommand{\nfig}[1]{Figure~\ref{#1}}
\newcommand{\speed}[1]{#1 km~s${}^{-1}$}

\begin{document}

\title{On a solar blowout jet: driven mechanism and the formation of cool and hot components}
\correspondingauthor{Yuandeng Shen}
\email{ydshen@ynao.ac.cn}
\author{Yuandeng Shen}
\affiliation{Yunnan Observatories, Chinese Academy of Sciences,  Kunming, 650216, China}
\affiliation{State Key Laboratory of Space Weather, Chinese Academy of Sciences, Beijing 100190, China}
\affiliation{Key Laboratory of Solar Activity, National Astronomical Observatories of Chinese Academy of Science, Beijing 100012, China}
\affiliation{Center for Astronomical Mega-Science, Chinese Academy of Sciences, Beijing, 100012, China}
\author{Ying D. Liu}
\affiliation{State Key Laboratory of Space Weather, Chinese Academy of Sciences, Beijing 100190, China}
\affiliation{University of Chinese Academy of Sciences, Beijing 100049, China}
\author{Jiangtao Su}
\affiliation{Key Laboratory of Solar Activity, National Astronomical Observatories of Chinese Academy of Science, Beijing 100012, China}
\affiliation{University of Chinese Academy of Sciences, Beijing 100049, China}
\author{Zhining Qu}
\affiliation{School of Physics and Electronic Engineering, Sichuan University of Science \& Engineering, Zigong 643000, China}
\author{Zhanjun Tian}
\affiliation{Yunnan Observatories, Chinese Academy of Sciences,  Kunming, 650216, China}
\affiliation{University of Chinese Academy of Sciences, Beijing 100049, China}

\begin{abstract}
We present the observations of a blowout jet that experienced two distinct ejection stages. The first stage started from the emergence of a small positive magnetic polarity, which cancelled with the nearby negative magnetic field and caused the rising of a mini-filament and its confining loops. This further resulted in a small jet due to the magnetic reconnection between the rising confining loops and the overlying open field. The second ejection stage was mainly due to the successive removal of the confining field by the reconnection. Thus that the filament erupted and the erupting cool filament material directly combined with the hot jet originated form the reconnection region and therefore formed the cool and hot components of the blowout jet. During the two ejection stages, cool H$\alpha$ jets are also observed cospatial with their coronal counterparts, but their appearance times are earlier than the hot coronal jets a few minutes. Therefore, the hot coronal jets are possibly caused by the heating of the cool H$\alpha$ jets, or the rising of the reconnection height from chromosphere to the corona. The scenario that magnetic reconnection occurred between the confining loops and the overlying open loops are supported by many observational facts, including the bright patches on the both sides of the mini-filament, hot plasma blobs along the jet body, and periodic metric radio type \uppercase\expandafter{\romannumeral3} bursts at the very beginnings of the two stages. The evolution and characteristics of these features manifest the detailed non-linear process in the magnetic reconnection.

\end{abstract}

\section{Introduction}
Solar jets are collimated hot plasma flows along vertical or oblique magnetic field lines. They are ubiquitous in the solar atmosphere from the photosphere to the outer corona. Although researches on solar jets have a long history, many physical questions about their detailed magnetic structure and formation mechanism are still unclear. Previous observational studies have suggested that most solar jets are formed due to the magnetic reconnection between emerging bipoles and their ambient open magnetic field lines \citep[e.g.,][]{shibata94b, liu04,shen11,li12,chen12,lih15,lim16}. However, sometimes magnetic flux cancellation is also important to the generation of solar jets \citep[e.g.,][]{shen12,yangj12a,yangj16,adams14,panesar16}. Moreover, \cite{shen14} reported that large-scale coronal waves can also lead to coronal jet by disturbing the coronal magnetic field at the boundary of coronal holes. Solar jet or jet-like phenomena are always associated with micro-flares that transfer magnetic energy to heat and kinetic energy. Since the occurrence rate of jet or jet-like activities is very high in the solar atmosphere, some solar physicists believe in that solar jets could be a possible candidate source for heating the coronal plasma and accelerating the fast solar wind \citep[][]{innes97,shibata07,tian14}. In addition, observational studies also indicated that some jet activities can directly or indirectly result in large-scale magnetic reconfigurations such as coronal mass ejections \citep[CMEs;][]{wang98,liu05,liu08,shen12,liu15}, filament or loop eruptions \citep[e.g.,][]{jiang08,wang16,zheng16}, and corona waves \citep{zheng12,zheng13,su15}. Therefore, investigation the driving and evolution mechanisms of solar jets are very important in solar physics.

Currently, there are three main classification methods for solar jets. Firstly, solar jets can be divided into surges (H$\alpha$), Extreme Ultraviolet (EUV) jets (EUV), and X-ray jets (X-ray) according to different observing wavelengths. Secondly, according to their different driving mechanisms and morphology, solar jets are classified into the so-called ``anemone jet'' and the dubbed ``two-sided loop jet'' \citep{shibata94a,yokoyama95,jiang13,tian17}. They are thought to be formed by magnetic reconnection between emerging bipoles and the ambient oblique or horizontal magnetic field lines, respectively. It should be pointed out that the two-sided-loop jets can also be produced via the so-called tether-cutting reconnection \citep{moore01,chen14,chen16,xue17} between two adjacent filamentary threads \citep{yang16,tian17}. Thirdly,  \cite{moore10} performed a statistical analysis and claimed that there is a dichotomy of coronal jets according to their different eruption characteristics. The authors named them ``standard jet'' and ``blowout jet''. The former is similar to the standard reconnection picture for coronal jets, while the latter is often associated with erupting loops or twisted small filaments in the jet-base. So far there are many studies based on high-resolution observations which have confirmed the finding of blowout jets \citep[e.g.,][]{shen12,yangj12b,pucci13,kayshap13,adams14,young14,hong13,hong16,hong17,lix15,sterling16,lim16,zhang17,zhu17}, and a few observational studies further investigated the relationship between blowout jets and CMEs \citep[][]{hong11,shen12}. Especially, \cite{shen12} reported that a single blowout jet can lead to a jet-like and a simultaneous bubble-like CMEs. They proposed that the jet-like CME is resulted from the outward moving hot plasma produced in the magnetic reconnection between the emerging bipole and the ambient open field lines, while the bubble-like CME is caused by the eruption of the mini-filament confined by the emerging bipole. Furthermore, numerical simulation studies of coronal blowout jets are also preformed by many authors \citep[e.g.,][]{pariat09,he10,pariat10,archontis13,ni15,pariat15,pariat16,wyper16a,wyper16b,wyper17,szente17}. Here, we would like to point out that before the bringing forward conception of the blowout jet, a few observations had indicated that some jets or surges are tightly related to mini-filament or loop eruptions \citep[][]{chen09,nistico09}. Very recently, \cite{hong14} observed many jet-like mini-filament eruptions from coronal bright points. \cite{sterling15} further proposed that almost all coronal jets are originated from mini-filament eruptions, and they also suggested that standard and blowout jets are fundamentally the same phenomenon. If the eruption of a filament is failed or succeeded to escape the confining closed-field jet-base, the consequence of the eruption is to form a standard or blowout jet, respectively. These studies together suggest that filament eruptions are deeply involved in jet activities, and different solar eruption phenomenon might obey a universal eruption model \citep{shen12,wyper17}.

In recent years, a lot of new characteristics of solar jets have been discovered by using high temporal and high spatial resolution observations. For example, \cite{shen11} presented an unwinding polar coronal jet which exhibits intriguing bright helical fine structure winding the jet body and three distinct expansion phases in the lateral direction. The unwinding motion of the coronal jet is thought to be caused by the releasing of magnetic twist stored in the emerging bipole via magnetic reconnection between the bipole and the ambient open field lines. The measured released magnetic twists during the ejection is about 1.17 -- 2.55 turns, which is comparable to typical active region filaments \citep{yan14,yan15}. In addition, the three distinct expansion phases are possibly associated with the different stages in the nonlinear magnetic reconnection process, and this hypothesis has recently been confirmed by \cite{chen17}. So far, there are many studies which also reported the rotational motion of coronal jets \citep[e.g.,][]{zhang00,liu09,curdt11,curdt12,hong13,schmieder13,lee13, zhang14a,filippov15,moore15,li17a,li17b}, and \cite{chen12} even estimated the magnetic field strength of a rotational jet to be 15 to 3 Gauss from the jet-base to the top. To the best of our knowledge, the rotational motion is not a common characteristic for all coronal jets, but it is indeed frequently observed in coronal blowout jets. This is possibly due to the reason that blowout jets are often associated with twisted filament eruptions that transfer magnetic twist to the open fields and therefore drive the jet's rotation \citep{shibata86,canfield96}. Sometimes, coronal jets can occur repeatedly at the same position, they are often  associated with successive magnetic flux emergence and cancellation processes at the jet source region \citep{jiang07,yang11,jiang13,chen08, chifor08,lih15}, as well as moving magnetic features \citep[e.g.,][]{brooks07,chen15}. Using magnetic extrapolation method and analyzing the change of current in the jet source region, \cite{guo13} proposed that periodic magnetic reconnection can result in recurrent coronal jets as well. In addition, the oscillation of coronal jets are thought to be the evidence of propagating Alfv\'{e}n waves \citep[][]{nishizuka08,liu09,lee13,lee15}. 

An important characteristic of solar jets is that some coronal jets are consisted of both cool and hot plasma flows \citep[e.g.,][]{mulay17}, and the cool component is often delayed to the hot one. \cite{canfield96} reported that chromospheric H$\alpha$ surges are spatially adjacent to the corresponding coronal X-ray jets. \cite{jiang07} also observed the similar spatial relationship between surges and coronal jets, and they further reported that the start of H$\alpha$ surges are slightly delayed to the beginning of the corresponding coronal jets. Previous studies indicated that the delay time interval ranges from 2 to 15 minutes, and most authors explained this phenomenon as the cooling of the earlier and hotter coronal jet material \citep[e.g.,][]{schmieder94, alexander99, jiang07}. However, there are still a few other interpretations for the formation and the delay of the cool component. For example, \cite{nishizuka08} reproduced both cool and hot components in their simulation and they interpreted the delay of the jet's cool component  resulted from the different Alfv\'{e}n velocities in the cool (high-density) and hot (low-density) plasmas rather than the cooling effect, since they found that the delay time is much shorter than the cooling time of the hot plasma flow. \cite{yokoyama95} also generated the hot and cool plasma components in their simulation, and they explained that the cool component of coronal jets is formed by chromosphere cool plasmas that are carried up with expanding loops and accelerated by the tension force of disconnected field lines. \cite{chae99} found that the coronal jets are identified with bright jet-like features in the H$\alpha$ line center, and they claimed that their observation confirms the simulation result of \cite{yokoyama95}. In addition, \cite{lee13} also proposed that the cool component of the jet are caused by reconnection of emerging flux in the transition region or upper chromosphere. Recently, \cite{shen12} observed the appearance of both hot and cool components in a typical coronal blowout jet with stereoscopic high resolution observations, they observed similar characteristics reported in previous articles, such as the spatial relationship and the delay appearance of the cool component. They clearly showed that the cool component of the coronal jet is actually formed by the erupting small filament that is confined by the jet-base, while the hot one resulted from the heated plasma during the reconnection between the jet-base and the ambient open magnetic field lines.  So far, the formation mechanism of the cool component of coronal jets is still an open question, and more details about coronal jets can be found in a recent review article \citep{raouafi16}.

In this paper, we present an observational study on the formation of the cool and hot plasma structures and plasma blobs in a miniature coronal blowout jet, using the high temporal and high spatial resolution data provided by the New Vacuum Solar Telescope \citep[NVST;][]{liu14,xiang16} and the Atmospheric Imaging Assembly \citep[AIA;][]{lemen12} onboard the {\em Solar Dynamics Observatory} \citep[{\em SDO};][]{pesnell12}. The present event occurred on April 16, 2014 at the southern periphery region of NOAA active region 12035. We find that the cool component of the coronal jet is directly formed by the eruption of a mini-filament at the jet-base. In addition, successive hot plasma blobs are observed moving along the jet body, and metric type \uppercase\expandafter{\romannumeral3} radio burst are also detected during the initiation and violent ejection phases of the jet. These result provide evidence for supporting the scenario that the blowout jet was driven by the  magnetic reconnection process. The used instruments and data set are briefly introduced in Section 2. The analysis results are described in Section 3. Conclusions and discussions are highlighted in the last section.

\section{Observations}
The NVST is a new one-meter solar telescope operated by the Yunnan Observatories, which locates at the northeast side of Fuxian Lake, Yunnan, China. The primary goals of NVST are high resolution imaging, spectral observation, and measurements of the solar magnetic field. Currently, the NVST mainly observes the photosphere and chromosphere using TiO-band, G-band, and H$\alpha$ lines. We only use the H$\alpha$ line center observations in the present paper, which has a cadence of 12 s and a spatial resolution of 0.3\arcsec. Due to the influence of the Earth's turbulent atmosphere, the raw solar images taken by the NVST are reconstructed using high-resolution imaging algorithms \citep{liu14}. The AIA onboard {\em SDO} images the full-disk Sun using seven EUV and three UV wavelengths, which takes EUV (UV) images every 12 (24) seconds and with a pixel size of 0\arcsec.6. The line-of-sight (LOS) magnetograms taken by the Helioseismic and Magnetic Imager \citep[HMI;][]{schou12} onboard {\em SDO} are also used to analyze the magnetic field topology, which has a cadence of 45 seconds and a measurement precision of 10 Gauss. All images used in this paper are differentially rotated to the reference time of 07:30:00 UT. In addition, metric radio observations recored by the metric spectrometer of the Yunnan Observatories \citep[YNAO;][]{gao14a} are also used in the present paper to diagnose the detailed magnetic reconnection process, which work in the frequency range of 70 -- 700 MHz with a spectral resolution of 200 kHz, a time cadence of 90 ms, and high sensitivity less than 1 sfu.

\section{Results}
The present event occurred on 2014 April 16 at the southern periphery region of NOAA active region AR12035 that was close to the solar disk center. The ejection of the blowout jet was associated with the eruption of a mini-filament that has a projection length of 13 Mm. According to previous statistical results, the length of the mini-filament is much less than the average projection length of mini-filaments (19 Mm) \citep{wang00}.

\begin{figure*}[thbp]
\epsscale{0.8}
\plotone{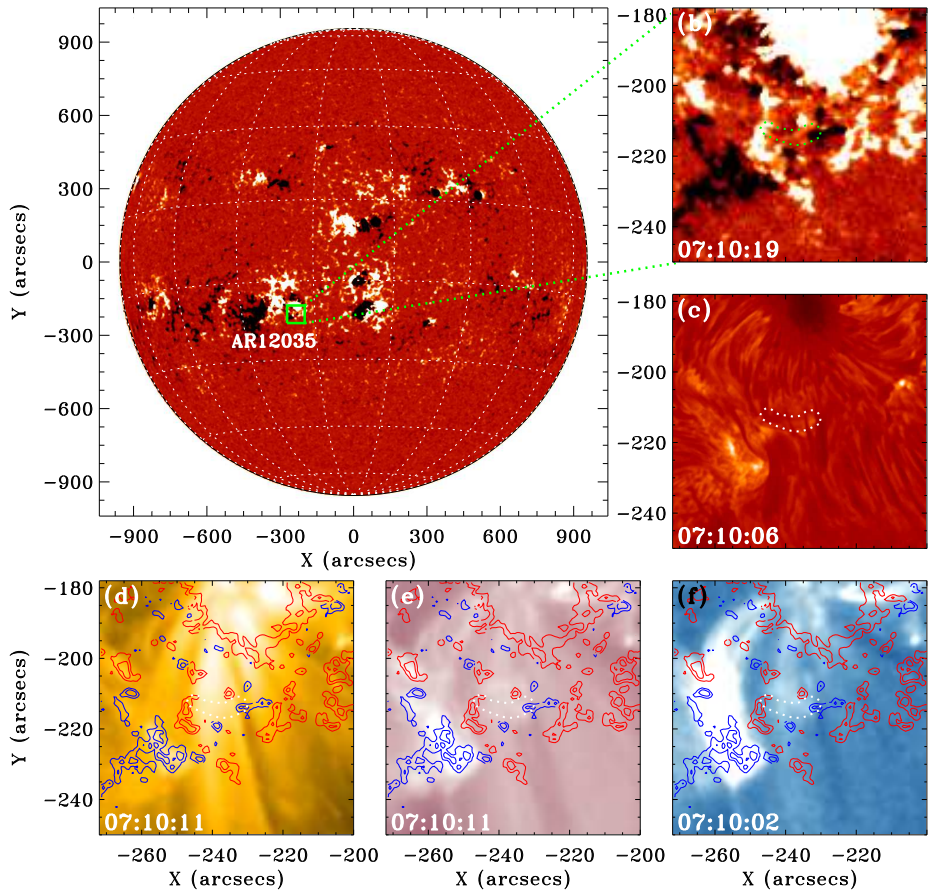}
\caption{An overview of the event before the eruption. (a) is a full-disk HMI LOS magnetogram, in which the white  and black patches represents the positive and negative magnetic polarities, respectively. The green box shows the eruption source region, and the details of which in other wavelengths are shown in the rest of panels. Panels (b) and (c) are the close view of the HMI LOS magnetogram and NVST H$\alpha$-center images overlaid with the outline contour sketch of the mini-filament. Panels (d), (e) and (f) are AIA 171, 211, and 335 images overlaid with the contours of the HMI LOS magnetic fields and the outline contour sketch of the mini-filament, in which the magnetic field contour levels are $\pm 50, \pm 100, \rm and \pm 200$, with red and blue colors for the positive and negative magnetic polarities.
\label{fig1}}
\end{figure*}

An overview of the eruption source region before the start of the jet is shown in \nfig{fig1}. It can be seen that the source region located at the southern periphery of the leading sunspot of active region AR12035, which is close to the disk center of the Sun. The magnetic field of the source region is a mixed polarity region that is composed of many small positive and negative polarities (see \nfig{fig1} (b)). A mini-filament can be identified from the H$\alpha$ image in the source region, and the outline contour sketch is overlaid on different wavelength images (see the white dotted contour in \nfig{fig1} (b) -- (f)). At the same time, the contours of the HMI LOS magnetic field are also overlaid on the EUV observations to show the position relation between the filament and the surrounding magnetic field. From \nfig{fig1}, the filament can be clearly identified in the NVST H$\alpha$ line-center image, but it is hard to find the counterpart in the EUV images. It should be noted that the filament can not be observed in the AIA 171 \AA\ image before the ejection (\nfig{fig1} (d)), since it was covered by a bunch of coronal loop rooted in the leading sunspot of NOAA active region AR12035.

\begin{figure*}[thbp]
\epsscale{0.8}
\plotone{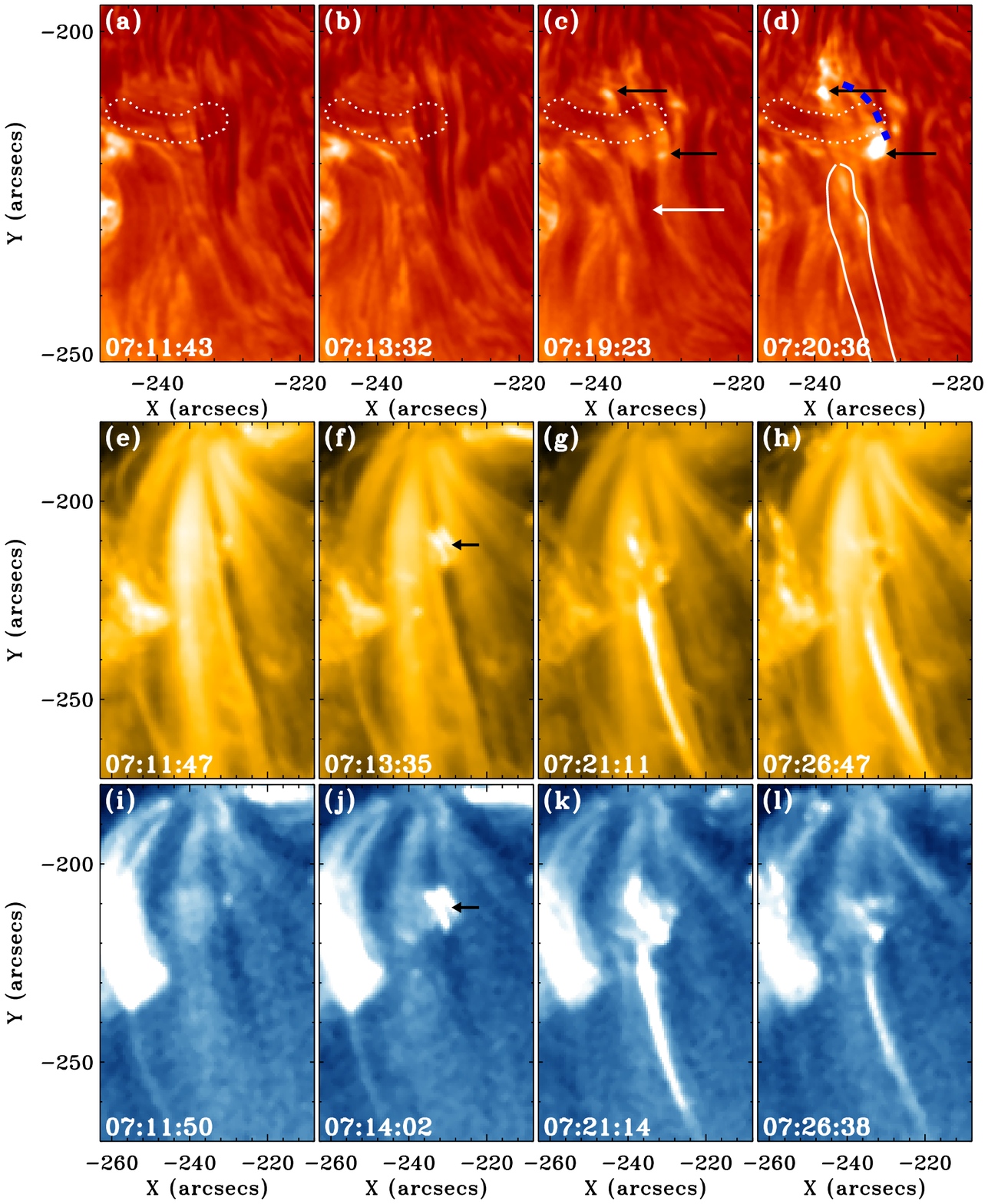}
\caption{The evolution of the first ejection stage of the blowout jet. (a) -- (d) are NVST H$\alpha$-center images; (e) -- (h) and (i) -- (l) are AIA 171 and 335 \AA images, respectively. The mini-filament is highlighted in the H$\alpha$ images with dotted contours. In panel (c), the two black arrows point to the pair of bright patches on the both sides of the mini-filament, while the white arrow points to the dark H$\alpha$ jet body. The arrows in panels (f) and (j) indicate the bright loop-like feature, whose position is overlaid in panel (d) as a blue dashed curve. The outline contour sketch of the coronal jet determined from the AIA 171 \AA image at 07:21:11UT is overlaid in panel (d). Animations are available in the online journal.
\label{fig2}}
\end{figure*}

The ejection of the blowout jet underwent two distinct ejection stages. The first ejection stage is shown in \nfig{fig2} with the NVST H$\alpha$ line center ($10^{4}$ K), AIA 171 ($6.3 \times 10^{5}$ K), and 335 \AA\ ($2.5 \times 10^{6}$ K) images. The top row shows the NVST H$\alpha$ line-center images overlaid with the outline contour sketch of the mini-filament in each panel (white dotted contour). Before the ejection, one can see that there is a bunch of dark filamentary thread (in north-south direction) across the western end of the filament (see \nfig{fig2} (a) and (b)). At about 07:11:43 UT,  the dark filamentary thread started to move to the east. About 8 minutes later, a pair of brightening patches appeared on the both sides of the filament at about 07:19:23 UT (see the two black arrows in \nfig{fig2} (c)). In the meantime, a small dark jet is observed in H$\alpha$ observations, and it showed eastward whip-like motion during the eruption (see the white arrow in \nfig{fig2} (c)). About one minute later, both the pair of brightening patches and the body of the jet became brighter than before (see the arrows in \nfig{fig2} (d)). It is hard to distinguish the ejecting H$\alpha$ jet from a single statical image, but it is clear by seeing the online animation made from NVST H$\alpha$ time sequence of images. 

In the AIA 171 and 335 \AA\ observations, a small loop-like bright structure is observed at the crossing position of the filamentary thread and the mini-filament during the first ejection stage. It is interesting that this loop-like bright feature exactly connects the pair of brightening patches observed in the H$\alpha$ images (see the blue curve in \nfig{fig2} (d)). This suggests that the bright loop-like structure and the pair of brightening patches could be regarded as the manifestations of the magnetic reconnection between the filamentary thread and the underneath closed loops that confine the mini-filament from eruption. A few minutes after the appearance of the bright loop-like feature, a collimated bright coronal jet is observed along the path of the H$\alpha$ filamentary thread. The outline contour sketch of the coronal jet detected from the AIA 171 \AA\ image at 07:21:11 UT is overlaid on the H$\alpha$ image at 07:20:36 UT (see \nfig{fig2} (d)), and the result shows that the trajectory of the H$\alpha$ jet and the bright coronal jet are the same. In addition, it is also found that the beginning of the H$\alpha$ jet precedes the appearance of its coronal counterpart about 2 minutes, which is inconsistent with previous observations that the cool component of solar jets delay the hot one a few minutes \citep[e.g.,][]{schmieder94, alexander99, jiang07}. However, the spatial relationship between the two jet components is in agreement with \cite{chae99}. It should be emphasized that \nfig{fig2} only shows some key snapshots of the ejection process. For detailed evolution process of the jet, one can see the associated animations available in the online journal.

\begin{figure*}[thbp]
\epsscale{0.8}
\plotone{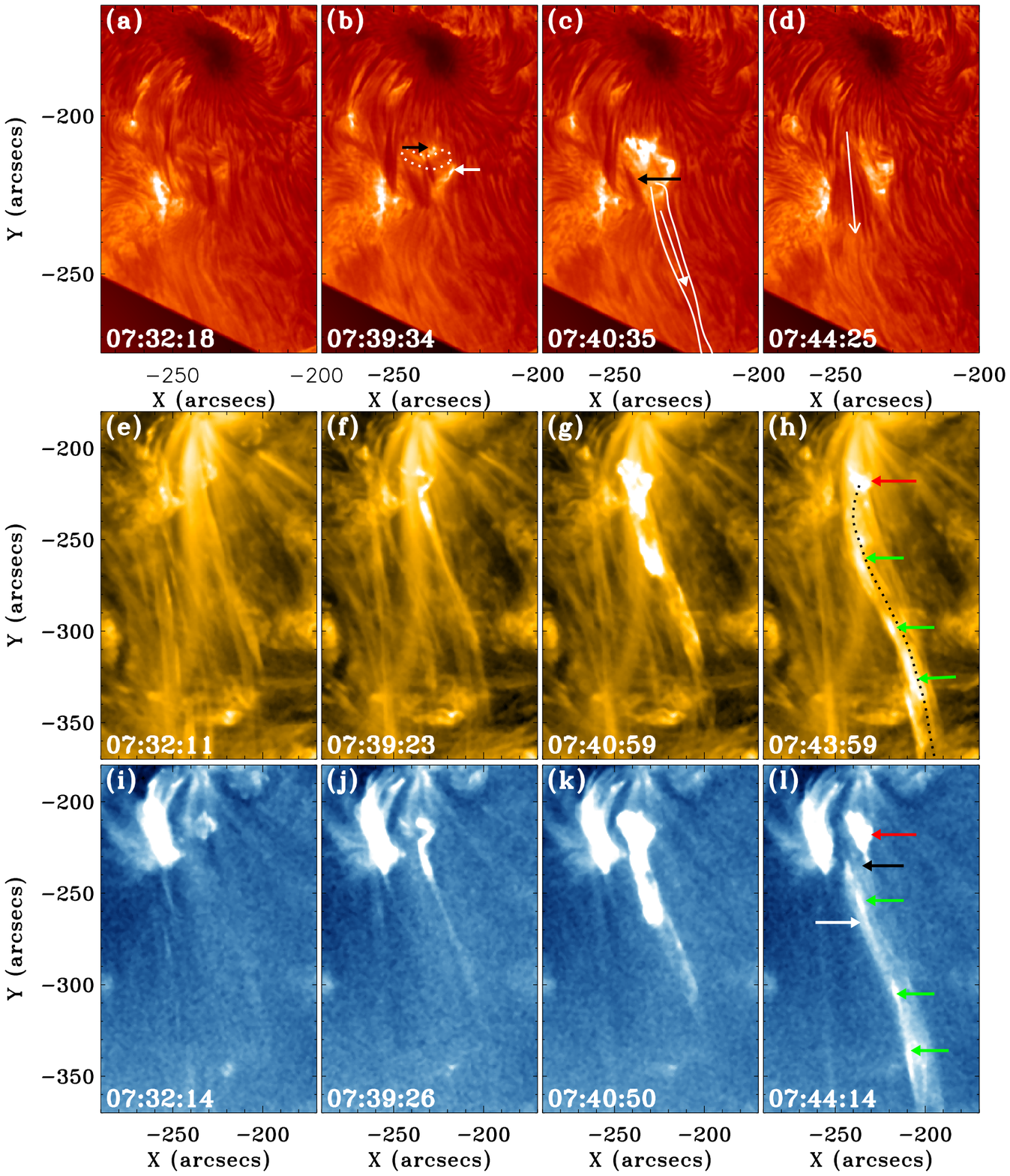}
\caption{The evolution of the second ejection stage of the blowout jet. (a) -- (d) are NVST H$\alpha$-center images; (e) -- (h) and (i) -- (l) are AIA 171 and 335 \AA images, respectively. The two arrows in panel (b) point to the bright patches on the both sides of the mini-filament which outlined by the white dotted contour. The arrow in panel (c) points to the erupting mini-filament, while the long white arrows in panels (d) and (d) indicate the direction of the H$\alpha$ jet. The white contour in panel (c) is the outline contour sketch of the coronal jet determined from the AIA 171 \AA image at 07:39:23 UT. The dotted curve in panel (h) shows the path used to obtain time-distance diagrams shown in \nfig{fig4}. The black and white arrows in panel (l) indicate the cool and hot components of the jet.  The red arrows in panels (h) and (i)  indicate the bright loop structure at the jet-base, and the green arrows point to the moving bright plasma blobs.
\label{fig3}}
\end{figure*}

\begin{figure*}[thbp]
\epsscale{0.7}
\plotone{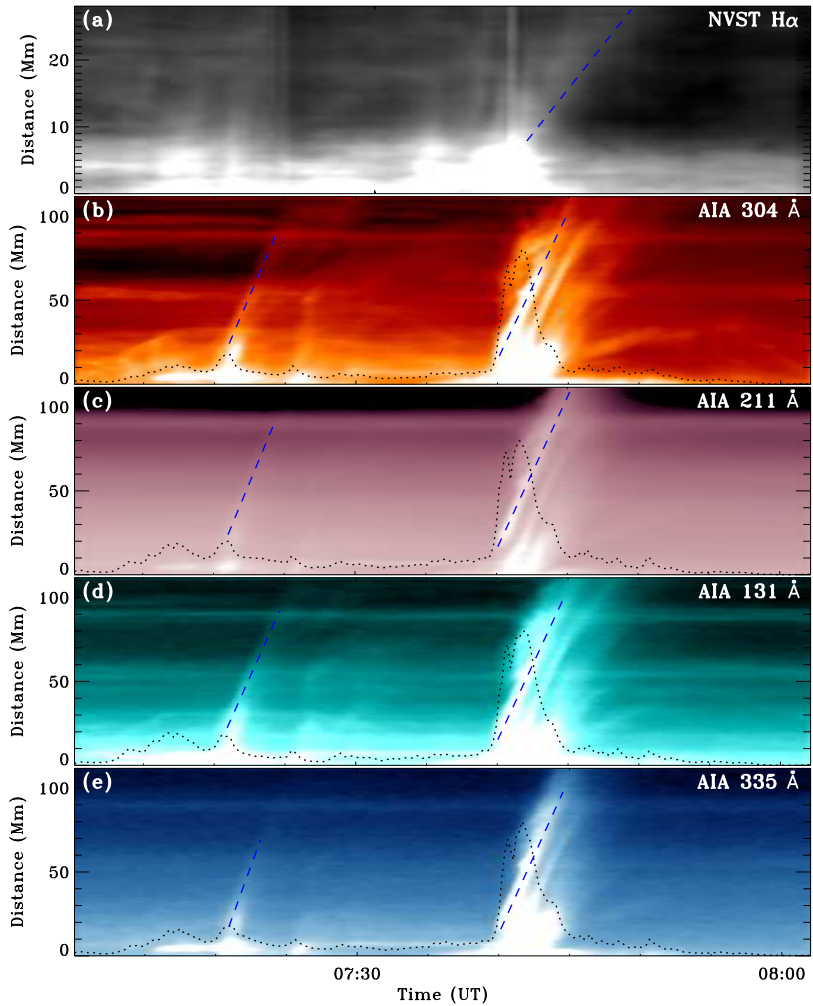}
\caption{Time-distance diagrams show the kinematics of the ejection blowout jet. (a) is made from the NVST H$\alpha$ observations, while (b) -- (d) are made from AIA 304, 211, 131, and 335 \AA base-difference images along the ejection direction as shown by the black dotted curve shown in \nfig{fig3} (h), respectively. The dotted lines are linear fit to the paths of the jets, and the corresponding wavelengths are also plotted on the top-right of each panel. The black dotted curve in panels (b) -- (e) show the normalized corresponding intensity lightcurves of the eruption source region.
\label{fig4}}
\end{figure*}

\begin{figure*}[thbp]
\epsscale{0.8}
\plotone{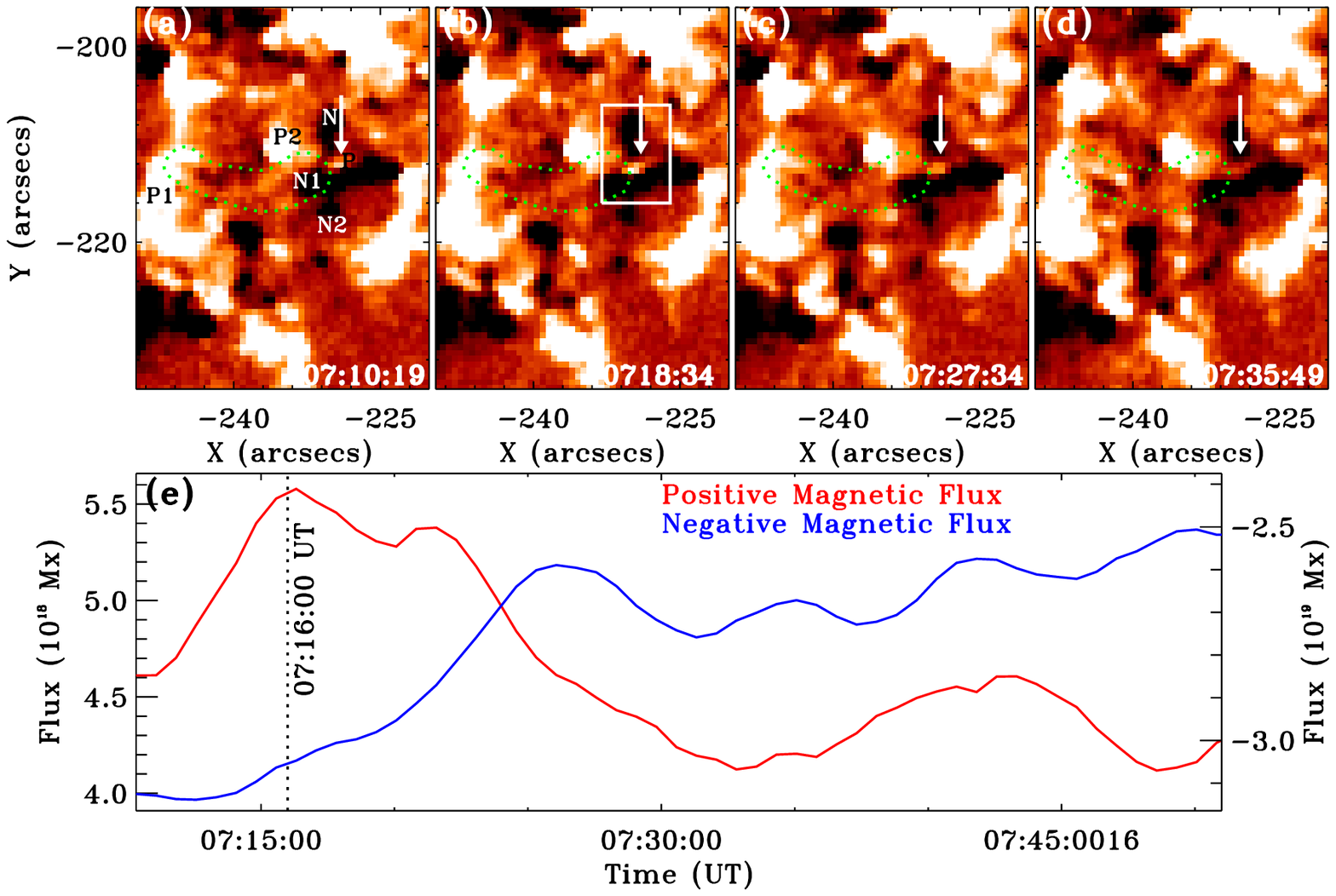}
\caption{The detailed evolution of the magnetic field of the eruption source region. (a) -- (d) are HMI LOS magnetograms. The green dotted curves overlaid on the magnetograms indicate the location of mini-filament, and the vertical white arrow points to a small positive magnetic polarity. The symbols ``P'', ``N'', ``P1'', ``N1'', ``P2'', and ``N2'' indicate the magnetic polarities involved in the present event. Panel (e) shows the variations of the positive (red) and negative (blue) magnetic fluxes in the white box region as shown in panel (b). The vertical dotted line in panel (e) indicates the start time of the magnetic cancellation.
\label{fig5}}
\end{figure*}

Following on the end of the first ejection stage, the blowout jet started a more violent ejection stage that showed interesting cool and hot plasma flows adjacent to each other and bright plasma blobs along the jet axis. Around the start of the second ejection stage, a pair of H$\alpha$ brightening patches appeared on the both sides of the filament as in the first ejection stage (see the two arrows in \nfig{fig3} (b)), then a relative larger H$\alpha$ jet formed and ejected to the south direction (see the bright long arrows in \nfig{fig3} (c) and (d)). In the meantime, the H$\alpha$ jet showed an eastward whip-like motion. This result is in agreement with the scenario that the H$\alpha$ jet is caused by the eruption of the mini-filament. In AIA 171 and 335 \AA\ EUV observations, a pair of bright patches and a bright thin coronal jet are observed at the jet-base and along the path of the H$\alpha$ jet, respectively (see panels (f) and (j) in \nfig{fig3}), and the appearance of the coronal jet delayed the H$\alpha$ jet a few minutes. The outline contour sketch of the EUV jet determined from the AIA 171 \AA\ image at 07:39:23 UT is overlaid on the H$\alpha$ image at 07:40:35 UT, and the result indicates that the H$\alpha$ jet is also cospatial with its coronal counterpart, in agreement with the first ejection stage. Around 07:40:50 UT, the erupting cool filament material merged into the previous existing bright hot jet and therefore formed the cool (dark) and hot (bright) components of the jet (see the black and white arrow in \nfig{fig3}(i)). The observational results indicate that the cool component of the jet is directly formed by the erupting filament. This is consistent with the results reported in \cite{shen12}, but is inconsistent with previous studies in which the cool component of jets are thought to be formed by the cooling of the hot component or other mechanisms \citep[e.g.,][]{schmieder94,alexander99,jiang07,nishizuka08,yokoyama95,lee13}. An interesting phenomenon of the jet during the second ejection stage is the formation of many bright plasma blobs that move along the jet axis (see the green short arrows in panels (h) and (i) in \nfig{fig3}), which may manifest the magnetic reconnection process in the formation of the coronal blowout jet. Similar observations of plasma blob structures in coronal jets have been documented in a few studies \citep[e.g.,][]{ohyama98,takasao12,kumar13,zhang14b,zhang16}, and very recently \cite{ni17} reproduced such physical process in their simulation.

\begin{figure*}[thbp]
\epsscale{0.8}
\plotone{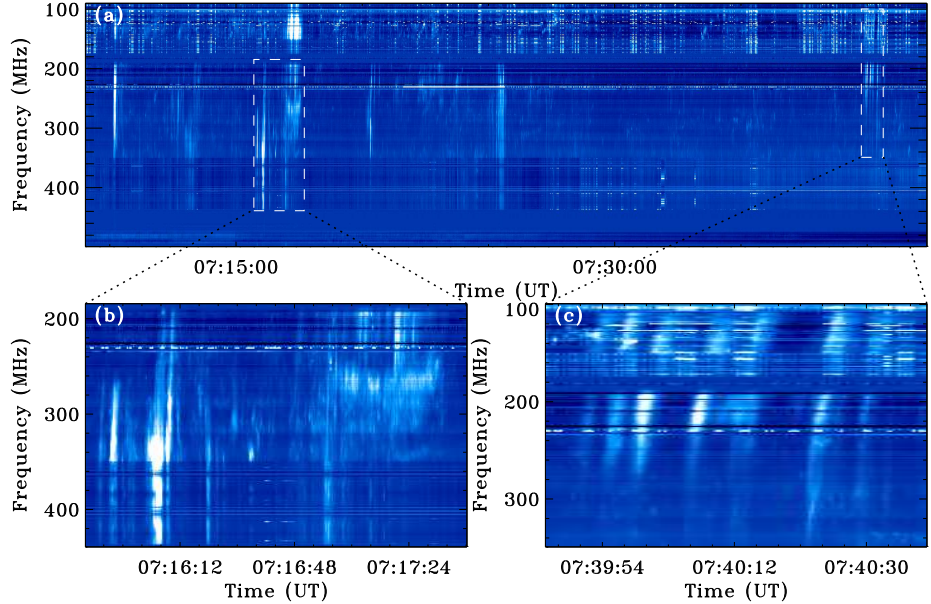}
\caption{Dynamic spectrum observed by the YNAO spectrometer. (a) shows the total of the right and left polarization spectrum maps from 07:09:02 UT to 07:42:22 UT in the frequency range of 90 -- 499 MHz. Panel (b) and (c) show the close up views of the two box region as shown in the top panel, and their time intervals are 120 and 52 second, respectively. The vertical bright ridges are the type \uppercase\expandafter{\romannumeral3} radio bursts. Note that the frequency ranges of panels (b) and (c) are 184 -- 439 and 99 -- 349 MHz, respectively.
\label{fig6}}
\end{figure*}

The kinematics analysis of the blowout jet is presented in \nfig{fig4} using time-distance diagrams made from NVST H$\alpha$, and AIA 304, 211, 131, and 335 \AA\ observations along the jet axis as dotted curve indicated in \nfig{fig3} (h), in which each time-distance diagram is obtained by composing the time sequence of intensity profiles along the jet axis. For each time-distance diagram, the abscissa and ordinate represent time and distance from the jet-base, respectively. The ejection speed of the jet can be obtained by measuring the slope of the inclined bright stripe which represent the ejecting jet. As shown in the figure, the two ejection stages of the blowout jet formed two bright stripes within about 40 minute. The jet formed in the first stage has a relatively simple structure and with an average speed of \speed{340}, while the one formed in the second stage is more complicated and with an average speed of \speed{300}. It can be seen that the jet structure in the second ejection stage in the AIA time-distance diagrams shows fine bright alternating dark stripes. This is caused by the ejecting plasma blobs along the jet axis rather than the cool (dark) and hot (bright) components as observed in the imaging observations. In the time-distance diagrams, each bright stripe represents a moving plasma blob. It is found that the moving speed of the blobs is in agreement with previous observational studies \citep{takasao12,kumar13,zhang14b,zhang16}. Here, the jet speeds are measured from the four AIA time-distance diagrams. The jets are hard to identified in the H$\alpha$ time-distance diagram (see \nfig{fig4} (a)). We trace the dark edge during the second stage of the jet to estimate the moving speed, which represent the top of the H$\alpha$ jet (see the dashed line in \nfig{fig4}(a)). It is obtained that the ejection speed is about \speed{50}, which is much less than that obtained from AIA observations.

\begin{figure*}[thbp]
\epsscale{0.8}
\plotone{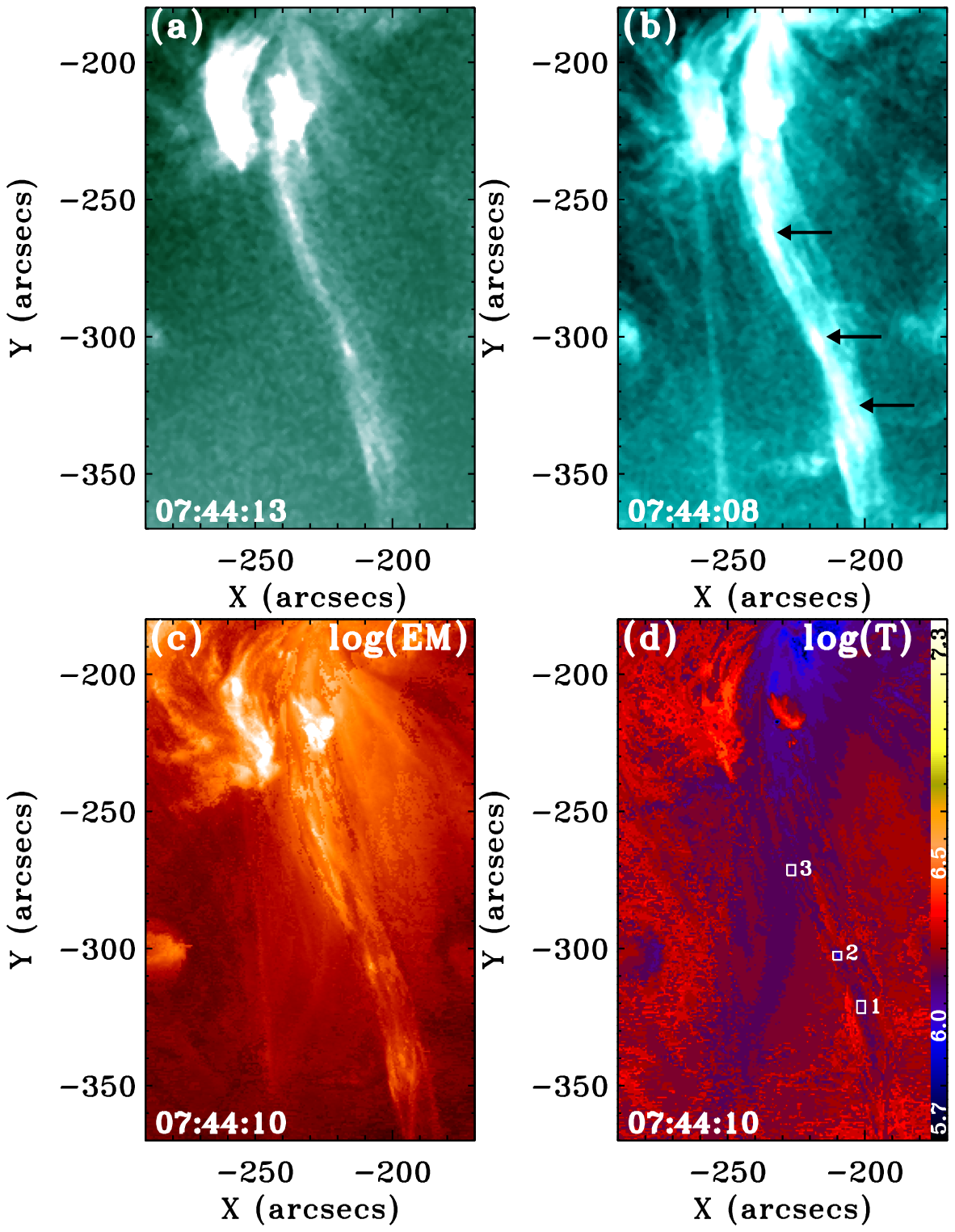}
\caption{DEM analysis of the plasma blobs at 07:44:10 UT. Panels (a) and (b) are AIA 94 and 131 \AA\ images, respectively. The moving plasma blobs are indicated with the three black arrows in panel (b). Panel (c) and (d) are the peak emission measure (cm$^{-5}$ K$^{-1}$) and temperature (MK) maps, respectively. The color bar on the right of panel (d) indicates the temperature variation range of the temperature map.
\label{fig7}}
\end{figure*}

\begin{deluxetable*}{cccccccc}[thbp]
\tablecaption{Temperature, density, and other parameters of the observed plasma blobs.
\label{tab1}}
\tablecolumns{8}
\tablenum{1}
\tablewidth{0pt}
\tablehead{
\colhead{Date/time} & \colhead{boxes} & \colhead{log($\rm T_{p}$)} & \colhead{log($\rm EM_{p}$)} &
\colhead{$\sigma_{T}$} & \colhead{TEM} & \colhead{L} & \colhead{$\rm n_{e}$} \\
\colhead{(UT)} & \colhead{ } & \colhead{(MK)} & \colhead{($\rm cm^{-5}\ K^{-1}$)} & \colhead{} &
\colhead{$\rm cm^{-5}$} & \colhead{(arcsec)} & \colhead{$\rm cm^{-3}$} }
\startdata
04/16/2014 07:44:10 & 1 & 6.18 & 20.92 & 0.27 & $6.06\times10^{28}$ & 3.2 & $1.61\times10^{10}$ \\
                                  & 2 & 6.16 & 21.05 & 0.33 & $1.88\times10^{29}$ & 5.5 & $2.16\times10^{10}$ \\
                                  & 3 & 6.18 & 21.21 & 0.16 & $3.52\times10^{29}$ & 5.8 & $2.89\times10^{10}$ \\
\enddata
\end{deluxetable*}

The evolution of the photospheric magnetic field in the eruption source region is displayed in \nfig{fig5} using the HMI LOS magnetograms, in which the white and black patches represent the positive and negative magnetic polarities, respectively. In the meantime, the outline contour sketch of the mini-filament is overlaid in each panel to show its position relation with the surrounding photospheric magnetic field. It can be seen that the mini-filament located on the magnetic polarity inversion line and with the both ends rooted in opposite polarities (P1 and N1). It is noted that a small southward moving positive magnetic polarity (P) near the west end of the filament interacted with the negative polarity where the western end of the filament rooted in. During the moving of the small positive polarity, whose area became smaller and smaller and eventually disappeared at about 07:35:49 UT (see the white arrow in the top row of \nfig{fig5}). This evolution process suggests that the small positive magnetic polarity cancelled with the nearby negative magnetic field (N1). We further measured the positive and negative magnetic fluxes in the white box region as shown in \nfig{fig5} (b), and the variation of the positive (red) and negative (blue) magnetic fluxes are plotted in \nfig{fig5} (e). It is found that the positive magnetic flux increases quickly during 07:11:00 UT to 07:16:00 UT and then it started a rapid decreasing phase. In the meantime, the negative magnetic flux decreases monotonously from 07:13:00 UT, but the decreasing speed obviously slowed down after 07:25:00 UT. The evolution process of the photospheric magnetic fluxes indicate that the magnetic cancellation was caused by the emerging and southward moving of the small positive magnetic polarity. Taking into account of the chromospheric and coronal eruption characteristics of the blowout jet, we propose that the initiation of the jet was caused by the emergence and cancellation of the small positive magnetic polarity at the western end of the mini-filament. The magnetic cancellation can first cause the disconnection of the western end of the mini-filament from the solar surface, and then the rising of the mini-filament and its overlying confining magnetic field. Eventually, the rising confining field of the mini-filament will reconnect with the overlying open magnetic field and therefore results in the observed solar jet in the first ejection stage.

This blowout jet was also recorded by the metric radio spectrometer of YNAO, and the dynamic spectrum is shown in \nfig{fig6}. From the radio spectrum shown in \nfig{fig6} (a), one can identify some small radio bursts at around 07:16:00 UT and 07:40:00 UT (see the two white box regions in \nfig{fig6} (a)). The two time slots well correspond to the very beginning phases of the two ejection stages of the blowout jet. The close up views of the two selected regions as shown by the two boxes in \nfig{fig6} (a) are plotted in \nfig{fig6} (b) and (c), respectively.  It is clear that there are many radio type \uppercase\expandafter{\romannumeral3} bursts at around 07:16:12 UT and radio type \uppercase\expandafter{\romannumeral3}-like bursts at around 07:40:12 UT that has a negative frequency drift. Since type \uppercase\expandafter{\romannumeral3} bursts represent the signature of propagating beams of nonthermal electrons in the solar atmosphere and are often associated with solar jets, they are thought to be resulted from the magnetic reconnection between closed and open magnetic fields \citep{chifor08,nitta08,kumar13,reid14,innes16,hong17}. Therefore, the detection of radio type \uppercase\expandafter{\romannumeral3} bursts at the very beginning of the first stage of the jet indicate the start of the magnetic reconnection process between the overlying open magnetic field and the confining loops of the mini-filament. Since the drifting radio type \uppercase\expandafter{\romannumeral3}-like bursts have a negative frequency drift rate, they suggest that the emission sources are moving from low to high altitude during the second ejection stage of the jet. By measuring the observed frequency of the radio type \uppercase\expandafter{\romannumeral3}-like bursts and the density model proposed in \cite{aschwanden95}, one can derive the propagation speed of the emission source \citep[e.g.,][]{gao14b,gao16}. Here, it is estimated that the upward moving speed of the radio emission source is about 30 Mm$^{-1}$, which is about 100 times of the ejection speed of the plasma blobs. With the wavelet analysis technique, it is obtained that the period of the radio type \uppercase\expandafter{\romannumeral3}-like bursts and the plasma blobs are about 4 and 90 second, respectively. Although the speeds and periods of the radio type \uppercase\expandafter{\romannumeral3}-like bursts show significant difference with those of the plasma blobs, both of the two phenomena are tightly related to the magnetic reconnection process that produces the observed jet.

\begin{figure*}[thbp]
\epsscale{0.8}
\plotone{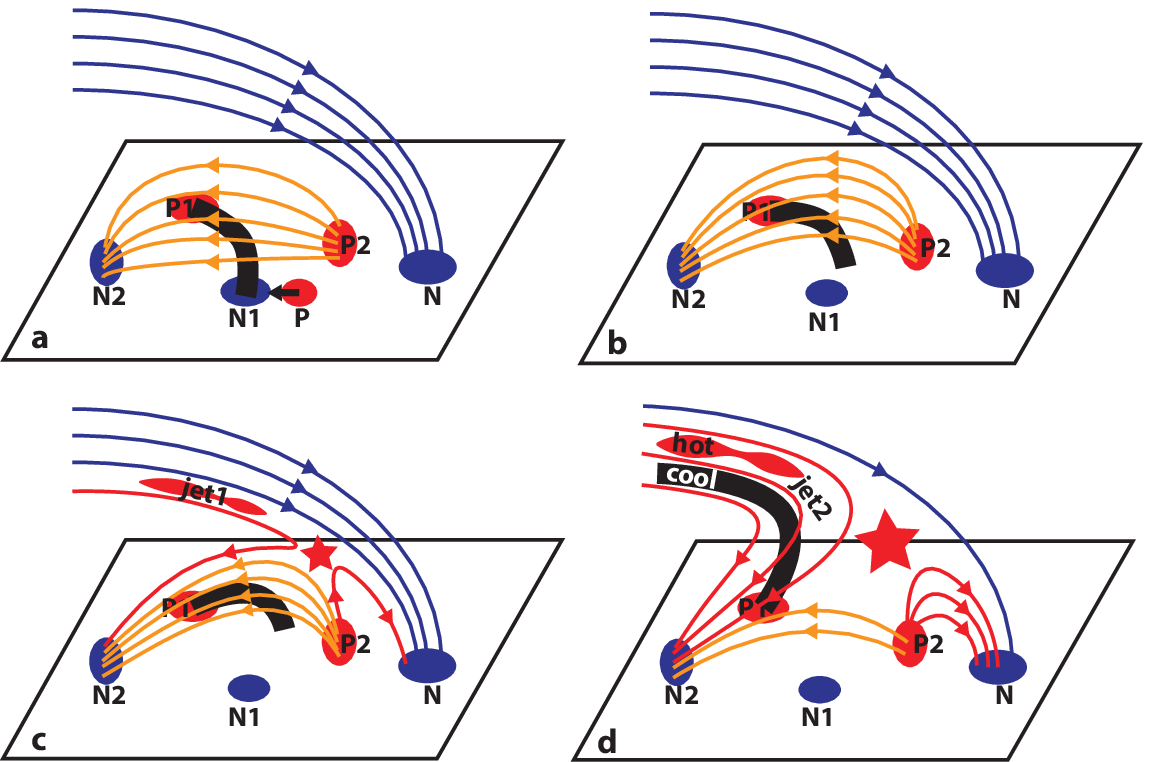}
\caption{A cartoon in analogy with the observation to illustrate the eruption process of the blowout jet. Panel (a) shows the topology of the magnetic field before the jet. Labels of P, P1, and P2 represent the positive magnetic polarities, while N, N1, and N2 represent the negative magnetic polarities. The mini-filament connects P1 and N1 and confined by a closed loop system connecting P2 and N2, and another open loop system rooted in N on the right. The black arrow indicate the moving direction of P.  Panel (b) shows the cancellation between P and N1 and the rising of the mini-filament. Panel (c) shows the magnetic reconnection between the closed and open loops and the formation of the jet during the first ejection stage. Panel (d) shows the eruption of of the mini-filament and the formation of the cool and hot components of the jet during the second ejection stage. The red lines show the reconnected magnetic field lines, and the red five-pointed star symbol indicate the reconnection position. The red sausage features in panels (c) and (d) indicate the hot plasma flow (jet) generated by the magnetic reconnection.
\label{fig8}}
\end{figure*}

To further analyze the physical property of the observed plasma blobs, high temperature wavelength filters of AIA 94 and 131 \AA\ are shown in \nfig{fig7} (a) and (b), respectively. The plasma blobs can be identified as bright patches along the jet axis as indicated by the arrows in \nfig{fig7} (b). The peak temperature and emission measure of the plasma blobs are analyzed with the differential emission measure (DEM) software available in the Solar SoftWare (SSW) package \citep{aschwanden13}. The code uses six coronal temperature wavelength filters of AIA (i.e., 94, 131, 171, 193, 211, and 335 \AA), and the parameters of the peak emission measure ($\rm EM_{p}$), the peak temperature ($\rm T_{p}$), and the temperature width sigma ($\rm \sigma_{T}$) can be obtained by fitting the DEM distribution function (Gaussian distribution) in each pixel. The emission measure and temperature maps of the jet are shown in \nfig{fig7} (c) and (d), respectively. It can be seen that the temperature of the eruption source region and positions of the plasma blobs along the jet are up to $\sim$ 6 MK. For comparison the physical parameters of the plasma blobs with previous observation \citep{kumar13}, we also calculate the total emission measure (TEM) in the selected regions using the formula $\rm \int DEM(T)dT$. Using this value we can estimate the densities ($\rm n_{e}$) of the plasma blobs by using the relation $n_{e} = \sqrt{\frac{\rm EM}{\rm L}}$ by assuming that the depth of the structure along the LOS is equal to its width \citep{cheng12}. Here, EM and L are the emission measure and the width of the blobs, and we assume the filling factor to be 1 as in \cite{kumar13}. Table 1 shows the all calculated parameters of the plasma blobs. It can be seen that the peak temperature of the plasma blobs varies from $\sim$1.4 to 1.5 MK, which is slightly lower than the value ($\sim$1.6 to 3.4 MK) presented in \citep{kumar13} and much lower than those ($\sim10$ MK) derived from X-ray observations \citep[e.g.,][]{ohyama98,nishizuka10}. The average density varies from $1.61\times10^{10}$, to $\rm 2.89\times10^{10}\ cm^{-3}$. It is found that the densities of the plasma blobs decrease with the increasing distance from the eruption source region of the jet, but their temperatures are almost the same. The decreasing density of the plasma blobs may suggest the dissipation process during their propagation.

\section{Interpretation}
In order to better understand the physics during the initiation and evolution of the blowout jet, a simple sketch is plotted in \nfig{fig8} to illustrate the detailed evolution process. It should be pointed out that only some representative magnetic field lines are plotted in the figure. \nfig{fig8} (a) shows the initial magnetic topology of the eruption source region, and the arrow on each curve indicates the direction of the magnetic field. Positive (P, P1, and P2) and negative (N, N1, and N2) polarities are analogy to the HMI LOS magnetogram as shown in \nfig{fig5} (a). According the observational results, the mini-filament connects magnetic polarities of P1 and N1 (black thick curve), and it is confined by a group of closed coronal loops that connects magnetic polarities of P2 and N2 (yellow). In addition, above the confining field of the mini-filament there is another group of open coronal loop that rooted in negative polarity N. In this magnetic topology, it is clear that a current sheet can be formed between the overlying open loops and the confining loop of the mini-filament, if the magnetic system subject to some external or internal disturbances.

The initiation of the blowout jet starts from the emergence of the small positive polarity P nearby the negative end of the mini-filament (N1), which approaches to N1 and caused the magnetic cancellation between polarities of P and N1. This will cause the disconnection of the mini-filament magnetic field from N1 and further lead to the rising of the mini-filament as well as the overlying confining coronal loops (\nfig{fig8} (b)). Due to the rising confining loops, a current sheet can be formed between the closed confining loops and the overlying open loops. The continue rising of the confining loops will trigger the magnetic reconnection in the current sheet, which not only produce the first ejection stage of the blowout jet but also the radio type \uppercase\expandafter{\romannumeral3} busts that manifest the upward moving beams of nonthermal electrons along the reconnected open magnetic loops (see \nfig{fig8} (c)). In the meantime, bright patches can be expect at polarities of N2, P2, and N due to the heating of the cool chromosphere material by the downward moving nonthermal particles along the reconnected field lines (red). In our observation, we do observe bright patches at N2 and P2, and the bright loop structure connecting P2 and N2 can also be explained as the heated confining loops due the magnetic reconnection. In our observation, the brightening patches and the radio type \uppercase\expandafter{\romannumeral3} busts provide evidence for the occurrence of magnetic reconnection between the closed and open magnetic fields. 

Another consequence of the magnetic reconnection between the closed and open field is the successive removal of the magnetic confinement that prevents the mini-filament from eruption. This will further accelerate the rising and the violent eruption of the mini-filament. As shown in \nfig{fig8} (d), the cool erupting filament material merges into the nearby hot jet from the reconnection region and therefore forms the cool and hot components of the blowout jet. The same as in the first stage, the pair of bright patches on the both sides of the mini-filament suggest the occurrence of magnetic reconnection between the rising closed confining loops and the overlying open loops. In addition, the observed plasma blobs are possibly due to the tearing instability during the turbulent magnetic reconnection, while the periodic radio type \uppercase\expandafter{\romannumeral3}-like busts during this stage are possibly caused by some nonlinear acceleration process of electrons during the magnetic reconnection.

\section{Conclussions \& Discussions}
Using high temporal and high spatial resolution observations taken by NVST, {\em SDO}, and the metric spectrometer of YNAO, we present the multi-wavelengths observation of a miniature solar blowout jet that occurred at the southern periphery of the preceding sunspot of NOAA active region AR12035. The blowout jet was associated with the eruption of a mini-filament that resides in the magnetic polarity inversion region and has a length of about 13 Mm. The ejection of the blowout jet can be divided into two stages. The first stage includes the initiation and ejection of a small jet, while the second one associates with the eruption of the mini-filament and the generation of the jet's cool and hot fine structures. It is measured that the ejection speeds of the jets during the first and the second stage are about \speed{340 and 300}, respectively. During the ejection, magnetic reconnection characteristics such as the bright patches on the both sides of the mini-filament, hot plasma blobs along the jet body, and periodic metric radio type \uppercase\expandafter{\romannumeral3} bursts around the beginning of the two ejection stages are observed, the appearance and evolution characteristics of these features manifest the detailed magnetic reconnection process between the closed confining loops of the mini-filament and the overlying open loops.

It is found that the initiation of the blowout jet was caused by the emergence of a small positive magnetic polarity, which cancelled with the negative magnetic field at western end of the mini-filament. The magnetic cancellation directly resulted in the rising of the mini-filament and its confining loops, which further caused the magnetic reconnection between the closed confining loops of the mini-filament and the overlying open loops. The heated plasma flow accelerated in the magnetic reconnection region formed the observed bright jet structure in the first stage. The observed bright patches on the both sides of the filament are possibly caused by the heating of the cool chromosphere material at the both ends of the closed reconnected loops by the downward nonthermal particles from the reconnection region. In the meantime, the reconnection also accelerate electrons moving along the open reconnected loops, which causes the observed metric radio type \uppercase\expandafter{\romannumeral3} bursts. During the first ejection stage, both dark (cool) H$\alpha$ and bright (hot) coronal jets are observed along the same trajectory, and they showed an obvious eastward whip-like motion. The appearance time of the cool H$\alpha$ jet was earlier than its hot coronal counterpart. This suggests that the height of the magnetic reconnection was possibly lower in the chromosphere, which first drives the H$\alpha$ jet and then the coronal counterpart. Another possibility is that the formation of the hot coronal jet is due to the heating of the cool H$\alpha$ counterpart in the chromosphere, since we do find that the cool H$\alpha$ jet became brighter than before at around the appearance of the hot coronal counterpart. \cite{chae99} also reported the observations of cospatial bright H$\alpha$ jets and hot coronal jets, and they proposed that the H$\alpha$ jets and the corresponding coronal jets are different kinds of plasma flows along different field lines but dynamically connected to each other.

The second ejection stage of blowout jet underwent a blowout process of the jet-base involving the mini-filament. The eruption of the mini-filament is due to two possible reasons: one is the magnetic cancellation at the western end of the mini-filament, which directly result in the disconnection of the mini-filament from the solar surface; The other is the successive removal of the magnetic confinement of the overlying loops since the magnetic reconnection during the first ejection stage. We observed many eruption characteristics during the second ejection stage of the blowout jet, including eastward whip-like motion of the jet body, two bright patches on the both sides of the filament connecting by a bright loop, periodic metric radio type \uppercase\expandafter{\romannumeral3}-like bursts with a period of 4 second at the beginning of the second ejection, and plasma blobs with a period of about 90 second along the jet body. We propose that all these observed eruption features are caused by the magnetic reconnection between the rising confining loops of the mini-filament and the overlying open loops, and the periodicity of the metric radio type \uppercase\expandafter{\romannumeral3}-like bursts and plasma blobs together manifest the detailed nonlinear  process in the magnetic reconnection. With the method of DEM, it is estimated that the temperature of the observed plasma blobs varies from $\sim$ 1.4 to 1.5 MK, which is slightly lower than the value presented in \cite{kumar13} and much lower than those derived from X-ray observations \citep[e.g.,][]{ohyama98,nishizuka10}. The estimated average density ranges from 1.61 $\rm 10^{10}$ to 2.89 $\rm 10^{10}\ cm^{-3}$, which shows a decreasing trend with the increasing distance from the eruption source region, but the temperature of the blobs at different distances are almost the same. This may suggest the dissipation process during the propagation of the blobs.

It is interesting that the erupting cool filament was adjacent with the previous existing bright jet body and therefore formed the observed cool (dark) and hot (bright) components of the blowout jet. It is measured that the appearance of the cool component delays the hot one about five minutes. Here, the formation mechanism of the cool and hot components of the blowout jet is consistent with the observations reported by \cite{shen12} but different with many other studies \citep[e.g.,][]{schmieder94, alexander99,jiang07,nishizuka08,yokoyama95,lee13}. However, temporal and spatial relationship between the two component are consistent with previous observational studies \citep[e.g.,][]{alexander99,jiang07}. It should be emphasized that there are two kinds of cool and hot components for solar jets. The first kind is that the two components are observed in cool (H$\alpha$) and hot (EUV or X-ray) lines, respectively. The other kind is that both the cool and the hot components are observed with hot coronal observations. In the present case, the cool and hot jets observed in the first ejection stage belongs to the first case, while the cool and hot components observed in the second ejection stage belongs to the second one.

Based on the observational results, a cartoon is proposed to interpret the ejection of the blowout jet. In our explanation, the cool jet component observed in the second ejection stage is directly resulted from the eruption of the mini-filament at the jet-base, while the hot one is the outward moving heated plasma flow generated in the magnetic reconnection. This interpretation is consistent with our previous observational result \citep{shen12}. The delayed appearance of the cool component is a natural consequence of the blowout jets in \cite{shen12} and the present event, because in both cases the start time of the mini-filament eruption is always after the magnetic reconnection that generates the hot plasma flow. The explanation of the formation and delay appearance of the cool component in blowout jet are different with various interpretations proposed in previous observational and simulation studies. For example, many authors believe that the cool component is formed by the cooling of the preceding hot component \citep[e.g.,][]{schmieder94,alexander99,jiang07}, or caused by reconnection of emerging  flux in the transition region or upper chromosphere \citep{lee13}. In addition, simulation works suggested that the cool component is formed by chromosphere cool plasmas that are carried up with expanding loops and accelerated by the tension force of disconnected field lines \citep{yokoyama95,chae99}, and \cite{nishizuka08} proposed that the delay of the cool component is due to the different Alfv\'{e}n velocities in the cool (high-density) and hot (low-density) plasma rather than the cooling effect. To better understand the formation mechanism of the cool component in solar jets, simulation works and more observational studies based on high temporal and high spatial multi-wavelength observations are needed in the future. 

\acknowledgments We thank the observations provided by NVST,  {\em SDO}, and the metric spectrometer of YNAO. We also thank the anonymous referee for his/her suggestions and comments that largely improve the quality of the present paper, and Dr. G. G in Yunnan observatories of Chinese Academy of Sciences for her help on the radio data analysis and explanation. This work is supported by the Natural Science Foundation of China (11403097,11633008,11773068), the Yunnan Science Foundation (2015FB191,2017FB006), the Specialized Research Fund for State Key Laboratories, the Open Research Program of the Key Laboratory of Solar Activity of Chinese Academy of Sciences (KLSA201601), the Youth Innovation Promotion Association (2014047) of Chinese Academy of Sciences.

\end{document}